\documentstyle[12pt,a4]{article}
\begin{document}
\newcommand{\wst}{~{}^{{}^{*}}\llap{$\it w$}}
\newcommand{\wdst}{~~~{}^{{}^{**}}\llap{$\it w$}}
\newcommand{\omegast}{~{}^{{}^{*}}\llap{$\omega$}}
\newcommand{\omegadst}{~~~{}^{{}^{**}}\llap{$\omega$}}
\newcommand{\must}{~{}^{{}^{*}}\llap{$\mu$}}
\newcommand{\mudst}{~{}^{{}^{**}}\llap{$\mu$~}}
\newcommand{\nust}{~{}^{{}^{*}}\llap{$\nu$}}
\newcommand{\nudst}{~{}^{{}^{**}}\llap{$\nu$~}}
\newcommand{\beq}{\begin{equation}}
\newcommand{\eeq}{\end{equation}}
\newcommand{\gfrc}[2]{\mbox{$ {\textstyle{\frac{#1}{#2} }\displaystyle}$}} 
\title {\large \bf Multidimensional Homogeneous Cosmological Models in
Wesson Theory of Gravitation }
\author{{\bf G S Khadekar} \thanks{Tel.91-0712-23946,
email:gkhadekar@yahoo.com} and {\bf Shilpa Samdurkar} \\
Department of Mathematics, Nagpur University \\
Mahatma Jyotiba Phule Educational Campus, Amravati Road  \\
Nagpur-440010 (INDIA) }
\maketitle
\begin{abstract}
Higher dimensional solutions are obtained for a homogeneous, spatially
isotropic cosmological model in Wesson theory of gravitation. Some
cosmological parameter are also calculated for this  model. 
\end{abstract}
\section{ Introduction}
Recently Wesson [1, 2] proposed a five dimensional theory of
 gravity where the rest masses varying with time. In this new theory,
 the space-time is no longer described by a four dimensional manifold
 but by a five dimensional space-time-mass (STM) Riemannian manifold
 where the mass plays the role of the fifth coordinate given by $ x^4
 =\frac{Gm}{c^2}(m= mass)$. The Einstein theory is recovered when the velocity
 $\frac{G}{c^2}\frac{dm}{dt} = 0,$ in other world, the mass is
 constant. In this regards the four dimensional theory of Einstein can
 be thought as embedded in a five dimensional STM theory.
\par This paper concerned with some possible exact solutions for a
 homogeneous, spatially isotropic n-dimensional $ (n > 5)$ cosmological
 model  in a matter free space. Although some exact vacuum solutions
 for five dimensional cosmological model have been worked out by Chi [3] and 
Chaterjee [4]. We have taken the matter density to be
 zero because otherwise  the higher dimensional field equations with
 both time and mass as a variable are too complicated to yield
 explicit solutions. Moreover for the particular case this empty space
 models provide instructive and transparent examples of various
 geometric possibilities. The paper ends with some comments and conclusions.
\par This type of problem has been discussed  in  STM theory of gravity by 
Chaterjee [4] for five dimensional case. However in our solutions, 
it is pointed out that at this stage, the implications of 
higher dimension solutions and is nterpretation are priliminary 
in nature because of some conceptual roblems associated with the new 
theory and because of the absence of
similar higher dimensional solutions in literature so for.
\subsection{Field Equations}
 The line element for a n-dimensional  homogeneous and
spatially isotropic  cosmological model is taken as \\
\begin{equation}
ds^2 = e^\nu dt^2-e^{\omega}\sum_{i=1}^{(n-2)}dx_{i}^2 + e^{\mu} dm^2
\end{equation}
where $ \mu,{\omega} $ and $\nu $ are the functions of time
and mass. Here the coordinate $x^0 = t, \: x^{1,2, \cdots, (n-2) }$  (space 
coordinate) and $x^{(n-1)} = m.$ For simplicity we have set the magnitudes of
 both c and G to unity. By applying this metric to the Einstein field equation
 $ G_{ij} = R_{ij} -\gfrc{1}{2} g_{ij} \, R = 0 $ with the assumption
$e^{\nu}= 1$  we get
\beq
G_{00} = -(n-2)(n-3) \; \frac{\dot \omega^2}{8}- (n-2) \: \frac{\dot \omega\dot 
\mu}{4} - (n-2) \, e^{-\mu}\left( \frac{\omegadst}{2}-\frac{\omegast \must}{4} + (n-1) \, \frac{\omegast^2 }{8}  \right)= 0
\eeq
\begin{eqnarray}
G_{11} = e^{\omega} \left(\frac{\ddot \mu}{2}  + \frac{\dot \mu^2}{4} 
\right) +  (n-3)e^{\omega}  \left(\frac{\ddot \omega }{2} +\frac{\dot{ 
\omega} \dot {\mu}}{4} +(n-2)\frac{{\ddot \omega^2} }{8} \right)
 \nonumber \\ +  (n-3)  e^{\omega- \mu} \left( \frac{\omegadst }{2} -
 \frac{ \omegast \must}{4} +  (n-2)  \frac{ \omegast^2}{8} \right)= 0  
\end{eqnarray}
 $$ G_{11} = G_{22} = G_{33} = \cdots = G_{(n-2)(n-2)} $$
\beq
G_{0(n-1)} = (n-2) \left( \frac{\dot{\omega}^*}{2 }+ 
\frac{\omegast \: \dot{ \omega}}{4}-\frac{\dot \mu \omegast}{4}\right) = 0 
\eeq
\beq
G_{(n-1)(n-1)} = -(n-2)(n-3) \frac{\omegast^2}{8} - (n-2) \;  
e^{\mu}\left(\frac{ \ddot \omega}{2} +(n-1)\frac{\dot{\omega^2}}{8}\right)=0 
\eeq
 where  a dot and star denote, respectively partial derivative with respect to time and mass.
\subsection{Solutions}
By solving equation (4) we get
\beq
 e^{\mu}= \alpha_{1}(m)\omegast^2  e^{\omega}
\eeq
where $\alpha_{1}(m)$ is an arbitrary function of mass. Using equation (6) in 
(5), we get   
\beq
\omegast^2\left(\frac{(n-3)}{8}+\alpha_{1}(m)e^{\omega}(\frac{\ddot \omega}{2} 
+ (n-1) \; \frac{\dot \omega^2}{8})\right) = 0
\eeq
since $\omegast^2$ is not equal to zero in higher dimensional metric
it follows that
\beq
\ddot \omega +\frac{(n-1)}{4} \:{\dot \omega^2} = 
-\frac{(n-3)}{4\alpha_{1}(m)}e^{-\omega} 
\eeq
which  yield the first integral 
\beq
\dot{\omega^2} = - \frac{1}{\alpha_{1}}e^{-\omega} + C_{1}(m) e^{ - 
\frac{(n-1)}{2}\omega} 
\eeq
Equation (9) can also be written as  
\beq
{\dot X^2} = -\frac{1}{ 4 \alpha_{1}} + \frac{C_{1}}{4} X^{-2k +2}
\eeq
where $ X = e^{\frac{\omega}{2}} $ and $ k= \frac{(n-1)}{2} $
\beq
or\; \int{\frac{dX}{\sqrt{C_{1} \alpha_{1} X^{2(1-k)}-1}}} = 
\int{\frac{dt}{2 \sqrt{\alpha_{1}}}}
\eeq
so that
\beq
 \frac{X \sqrt{1- \alpha_{1}\; C_{1} X^{3-n}}\; F_{1}[
\frac{1}{3-n},\;\frac{1}{2}, \; 1+\frac{1}{3-n},\; \alpha_{1} \; C_{1}X^{3-n}]}{\sqrt{\alpha_{1}\; C_{1}X^{3-n} -1}} = 
\frac{1}{2\sqrt{\alpha_{1}}} t + C_{2}
\eeq
where $\; F_{1}$ is a hypergeometric function  and $ C_{2} $ be the arbitrary function of mass. \\
For  $ C_{1} =0 $ in equation (9), we get
$$ e^{\omega} = \frac{ -t^2 + 2bt -b^2}{4\alpha_{1}} $$ 
\beq
 or \; \; e^{\omega} =-\alpha t^2 + \beta t + \gamma  
\eeq
where $ \alpha = \frac{1}{4\alpha_{1}}$,  $ \beta =
\frac{1}{2\alpha_{1}}$, $ \gamma = -\frac{b^2}{4\alpha_{1}} $  are
all functions of mass $ m $ only.\\ Equation (13) is similar to the
equation obtained by Chaterjee [4] for five dimensional case.
\par We can set $ \mu = 0 $ in the line element (1) instead of $ \nu =
0, $ and the resulting solution is
\beq
\frac{X \sqrt{1- A B X^{3-n}}\; F_{2}[
\frac{1}{3-n},\;\frac{1}{2}, \; 1+\frac{1}{3-n},\; A  B
X^{3-n}]}{\sqrt{ A B X^{3-n} -1}} = \frac{1}{2 \sqrt{A}}\; m + C
\eeq 
where $ F_{2} $ is hypergeometric function and $ A, B $ and  $ C $ are all
functions of time t only.
\subsection{Conclusion}
\par In this paper we have considered the the n-dimensional spatially
homogeneous and isotropic cosmological model in STM theory of
gravity. The main defect of the present model appears to be the arbitrary
nature of mass functions. However, as the role of mass in the STM
theory is not yet fully understood, neither in its geometrical
concepts nor in its physical realm. We think that this new exact
higher dimensional solutions together with cosmological
considerations should bring some additional information, and as such, 
they need to be further investigated. It is our hope that the higher 
dimensional solutions  presented here can be
 used as the starting point to investigate the behaviour of the rest
 of the particles in more realistic universe models.
\bibliographystyle{plain}

\end{document}